\documentclass[prl,showpacs,twocolumn,superscriptaddress,letterpaper]{revtex4}
\usepackage{amsmath}
\usepackage{amssymb}
\usepackage{graphicx}
\usepackage{xcolor}

\newcommand{\SHS}{Shastry-Sutherland }
\newcommand{\SCBO}{S\lowercase{r}C\lowercase{u}$_2$(BO$_3$)$_2$ }
\newcommand{\SCBOO}{S\lowercase{r}C\lowercase{u}$_2$(BO$_3$)$_2$}

\newcommand{\J}{$J$}
\newcommand{\Jp}{$J^{\prime}$}

\newcommand{\JJ}{$J$ }

\begin{document}
\title{Correlated decay of triplet excitations in the Shastry-Sutherland compound SrCu$_2$(BO$_3$)$_2$}

\author{M. E. Zayed}
\email[]{mohamed.zayed@qu.edu.qa}
\affiliation{Department of Mathematics, Statistics and Physics, College of Arts and Science, Qatar University, P.O. Box 2713 Doha, Qatar}
\affiliation{Laboratory for Quantum Magnetism, \'{E}cole Polytechnique F\'{e}d\'{e}rale de Lausanne (EPFL), CH-1015 Lausanne, Switzerland}
\affiliation{Laboratory for Neutron Scattering and Imaging, Paul Scherrer Institute, CH-5232 Villigen PSI, Switzerland}
\author{Ch. R\"uegg}
\affiliation{Laboratory for Neutron Scattering and Imaging, Paul Scherrer Institute, CH-5232 Villigen PSI, Switzerland}
\affiliation{DPMC-MaNEP, University of Geneva, CH-1211 Geneva, Switzerland}
\affiliation{London Centre for Nanotechnology and Department of Physics and Astronomy, University College London, UK} 
\author{Th. Str\"assle}
\affiliation{Laboratory for Neutron Scattering and Imaging, Paul Scherrer Institute, CH-5232 Villigen PSI, Switzerland}
\author{U.~Stuhr}
\affiliation{Laboratory for Neutron Scattering and Imaging, Paul Scherrer Institute, CH-5232 Villigen PSI, Switzerland}
\author{B. Roessli}
\affiliation{Laboratory for Neutron Scattering and Imaging, Paul Scherrer Institute, CH-5232 Villigen PSI, Switzerland}
\author{M. Ay}
\affiliation{Laboratory for Neutron Scattering and Imaging, Paul Scherrer Institute, CH-5232 Villigen PSI, Switzerland}
\author{J. Mesot}
\affiliation{Laboratory for Neutron Scattering and Imaging, Paul Scherrer Institute, CH-5232 Villigen PSI, Switzerland}

\affiliation{Laboratory for Neutron and Synchrotron Spectroscopy, \'{E}cole Polytechnique F\'{e}d\'{e}rale de Lausanne (EPFL), CH-1015 Lausanne, Switzerland}

 \affiliation{Laboratory for Solid State Physics, ETH Zurich, CH-8093 Zurich, Switzerland}

\author{P. Link}
\affiliation{Forschungsneutronenquelle Heinz Maier-Leibnitz (FRM-2), D-85747 Garching, Germany}

\author{ E. Pomjakushina}
\affiliation{Laboratory for Developments and Methods, Paul Scherrer Institute, CH-5232 Villigen PSI, Switzerland}

\author{M. Stingaciu}
\affiliation{Laboratory for Developments and Methods, Paul Scherrer Institute, CH-5232 Villigen PSI, Switzerland}

\author{K. Conder}
\affiliation{Laboratory for Developments and Methods, Paul Scherrer Institute, CH-5232 Villigen PSI, Switzerland}

\author{H. M. R$\o$nnow}
\affiliation{Laboratory for Quantum Magnetism, \'{E}cole Polytechnique F\'{e}d\'{e}rale de Lausanne (EPFL), CH-1015 Lausanne, Switzerland}

\date{\today}
\begin{abstract}
The temperature dependence of the gapped triplet excitations (triplons) in the 2D
Shastry-Sutherland quantum magnet \SCBO is studied by means of inelastic neutron scattering.
The excitation amplitude rapidly decreases as a function of temperature
while the integrated spectral weight can be explained by an
isolated dimer model up to 10~K. 
Analyzing this anomalous spectral line-shape in terms of damped
harmonic oscillators shows that the observed damping is due to a
two-component process: one component remains sharp and resolution limited
while the second broadens.
We explain the underlying mechanism through  a
simple yet quantitatively accurate model of correlated decay of triplons: 
an excited triplon is long-lived if no thermally populated triplons are near-by but decays quickly if there are.
The phenomenon is a direct consequence of frustration induced triplon localization in the Shastry--Sutherland lattice.
\end{abstract}


\maketitle

%%%
Quantum spin systems display a wide range of intriguing many-body quantum effects. A particularly active field is the study of interacting dimers systems. Two antiferromagnetically coupled spins forming a dimer have a singlet ground state with an energy gap to excited triplet states. In extended system where dimers couple to each other, the ground state often remains a spin singlet and gapped. The excitations are known as \emph{triplons}, and can be described in terms of quasi-particles as hardcore bosons~\cite{Gimarchi08,Sachdev90,Knetter01,Trebst00}. Due to coupling between dimers, triplons usually become mobile and can hop to neighboring dimer sites. For particular cases, the thermodynamic finite temperature behavior of the mobility of these hardcore bosons has been treated using statistical models reproducing triplon band renormalisation and damping observed in experiments~\cite{Ruegg05,Zheludev08,Tennant03,Tennant12,Castro12}. In general the increase of the thermal population of bosons, produces an increased repulsion and a reduced mobility, which is observed as a reduction of the dispersion band width. The theoretical treatment of finite temperature damping, reflecting the life time of the boson, and the related spectral functions remains an ongoing challenge~\cite{Essler08,Normand11,Fischer10,Jensen14}. 

The compound SrCu$_2$(BO$_3$)$_2$  constitutes an important example for testing our understanding of quantum spin systems as it is a close realization of the frustrated but \lq{}exactly solvable\rq{} 2D Shastry-Sutherland model~\cite{Kageyama99a_s,Shastry81}(see Ref.~\cite{Miyahara03R,KageyamaRR05} for a review). In \SCBO triplons are prevented from hopping already at zero temperature by frustrated inter-dimer interactions. Despite the strong (frustrated) coupling, the triplon dispersion is very shallow. Theoretical studies show that hopping is allowed only from the 6$^{th}$ order in the inter- to intra-dimer coupling ratio \Jp/\JJ or in the presence of  Dzyaloshinskii-Moriya (DM) terms. It has however been shown that multiple triplons forming a bound state are more mobile than a single triplon due to correlated hopping processes, appearing already in second order in \Jp/\J~\cite{Momoi00a,Miyahara03R}.  In high magnetic fields, triplons in \SCBO crystallize in regular patterns at fractional values of the saturation magnetization causing plateaus in the magnetization curve~\cite{Kageyama99a_s,Kodama02_short, Dorier08}.

The finite temperature properties of \SCBO have been noticed to be unusual but their understanding has remained elusive~\cite{Kageyama00a_s,Gaulin04_s,Shawish06,Haravifard06_s,Lemmens00a_short,Nojiri00b_short,Levy08,Zayed14}. Magnetization plateaus, for instance, are destroyed at small finite temperatures (about 1~K). In zero field the intensity of the triplet excitation measured by inelastic neutron scattering (INS) decays faster than expected for a dimer system with a gap $\Delta$=35~K. This implies that very small amounts of thermal energy dramatically modify the spin correlations of the system. \SCBO has been proposed in the context of valence bond crystals~\cite{Lauchli02}, resonating valence bond states~\cite{Takushima01,Koga00a,Miyahara03R}, superfluid, supersolid~\cite{Momoi00a,Schmidt08}, and (upon doping) superconducting phases~\cite{Shastry02,Kimura04,Yang08}. The understanding of its finite temperature properties is therefore of crucial importance.   

Supported by detailed INS measurements, we develop here a new scenario for interacting quasi-particle bosons (\emph{e.g.} triplons) on a 2D lattice, where the damping of a boson critically depends on the existence of another boson in its vicinity. This scenario explains the unusual low temperature behavior observed in \SCBOO.  The frustrated inter-dimer interaction causes the bosons to remain well localized, but on the other hand it is strong enough to cause a triplet to polarize neighboring singlets and thereby creates an extended boson covering several dimers.  It is then observed that the bosons decay when they overlap with one another. The combined effect of the strong inter-dimer coupling and of the frustration allows in \SCBO a unique real space insight of the boson interactions in strongly correlated systems.  
  
\begin{figure}[h]
\includegraphics[width=\columnwidth]{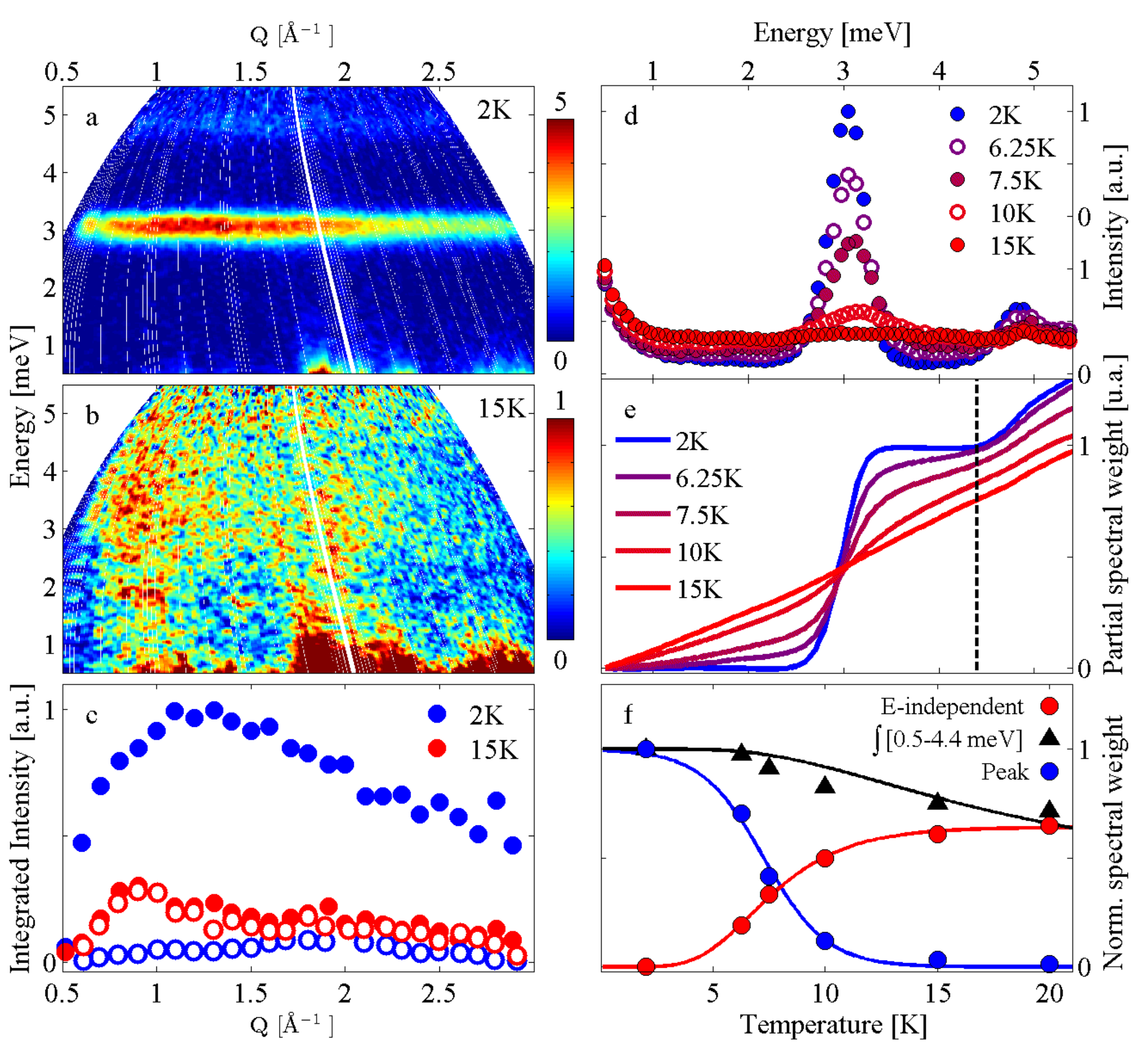}
\caption{(color online) Temperature dependence of the excitation spectrum of
\SCBOO, measured by neutron time of flight spectroscopy on the FOCUS spectrometer at SINQ on a powder sample. (a-b) Energy-momentum (E,Q) map at 2~K and 15~K. 
(c) Q dependence of the integrated intensity around the gap energy (E=[2.8-3.2]~meV, full symbols) and away from the gap energy (E=[3.8-4.2~meV] and E=[1.5-2.2~meV] combined, open symbols). (d) Q-integrated excitation spectra. (e) Partial spectral weight in the [0.5~meV-E] range as function of E. The dashed line at 4.4~meV marks the onset of the two-triplet bound state. (f) Scaled spectral weights from Gaussian fits to (d) showing the peak intensity and the energy-independent component, as well as the
integrated spectral weight in the [0.5-4.4]~meV range. The black line is the  $\Delta$ =35~K  isolated singlet population (see text), color lines are guides to the eye.}
\end{figure}

Powder samples of SrCu$_2$(BO$_3$)$_2$ were synthezised by solid state reaction. Single crystals were grown by traveling solvent floating zone technique. For measurements on the FOCUS time-of-flight spectrometer at SINQ, Switzerland, 12~g of powder sample was placed in an aluminium can, and data corrected for the background measured from an identical empty can. The incident neutron energy was 7.0~meV. Zero field measurements on single crystals were performed at the TASP triple-axis-spectrometer at SINQ using 2 crystals of total mass 8~g coaligned with $a$ and $c$ axes -- for Q=(1.5,0.5,0)  r.l.u. (reciprocal lattice units)-- and $a$ and $b$ axes -- for Q=(1.5,0,0) and Q=(2,0,0)-- in the horizontal scattering plane. Final neutron energy was 3.0~meV, giving an energy resolution (FWHM at 3~meV) of 0.19~meV. Magnetic field dependent measurements were performed on the PANDA spectrometer at FRM-2, Germany, using a 3~g single crystal aligned with $a$ and $b$ axes in the scattering plane. Magnetic field was applied along the vertical $c$ axis using a split coil cryomagnet. Final neutron energy was 5.0~meV, giving an energy resolution of 0.24~meV.

The ground state of SrCu$_2$(BO$_3$)$_2$ is well described by a product of singlets having a $\Delta$=35~K gap (corresponding to 3.0~meV) to very localized excited triplet states. A two-triplet bound state forms around 5~meV.

This is clearly observed at $T=$2~K in Fig.~1a where the INS  intensity  is shown for a powder sample as a function of momentum Q and energy E. The INS intensity is proportional to the spectral density and hence its integral will provide the spectral weight. At 15~K $<\Delta/2$  (Fig.~1b), however, the triplet excitation has disappeared. We present in Fig.~1c the Q dependence of the integrated INS intensities at different energies. At the lowest temperature, the spectral weight at the gap energy (closed symbols) has a maximum corresponding to the \SCBO dimer structure factor followed by a decrease due to the magnetic form factor of the Cu$^{2+}$ ion. The spectral weight away from the gap energy (open symbols) is negligible at 2~K, but as temperature increases it acquires more structure, and at 15~K, it eventually becomes identical to the Q dependence observed at 3~meV. 

The dispersionless nature of the excitation justifies looking at the Q-integrated scattering. The observed Q-integrated intensity (Fig.~1d) starts to be strongly suppressed above $T=$5~K$~\ll\Delta$ and turns into a flat
continuum around 15~K. The excitation broadens as the peak
intensity reduces, and its center moves slightly towards higher
energies.
We point out in Fig.~1e an interesting behavior of the partial spectral integral (i.e. spectral weight as function of the upper integration limit). It
demonstrates that the intensity is redistributed symmetrically to higher and
lower energies since the same spectral weight is recovered for all
temperatures at the gap energy. This behavior is robust upon reducing the
lower integration limit.\\ 

Gaussian fits to the Q-integrated spectra quantify the reduction of the spectral weight concentrated in a sharp peak and show an associated increase of a spectral component that is nearly energy independent (Fig.~1f). Since the neutron excites a dimer from a singlet state to a triplet state, the INS intensity is proportional to the
number of singlets. We thus compare the singlet population $n_s(T)=(1+3e^{-\Delta/T})^{-1}$ with a gap $\Delta$=35~K to the
spectral weight in the [0.5-4.4]~meV range. We find that this total spectral weight as opposed to that of the sharp peak remains consistent with the calculated singlet population. It thus appears that the total number of singlets follows the expectation for a gap $\Delta$=35~K, but that triplets become overdamped already at $T\ll\Delta$, removing spectral weight from the coherent quasi-particle peak.\\
\begin{figure}[h]
\includegraphics[width=\columnwidth]{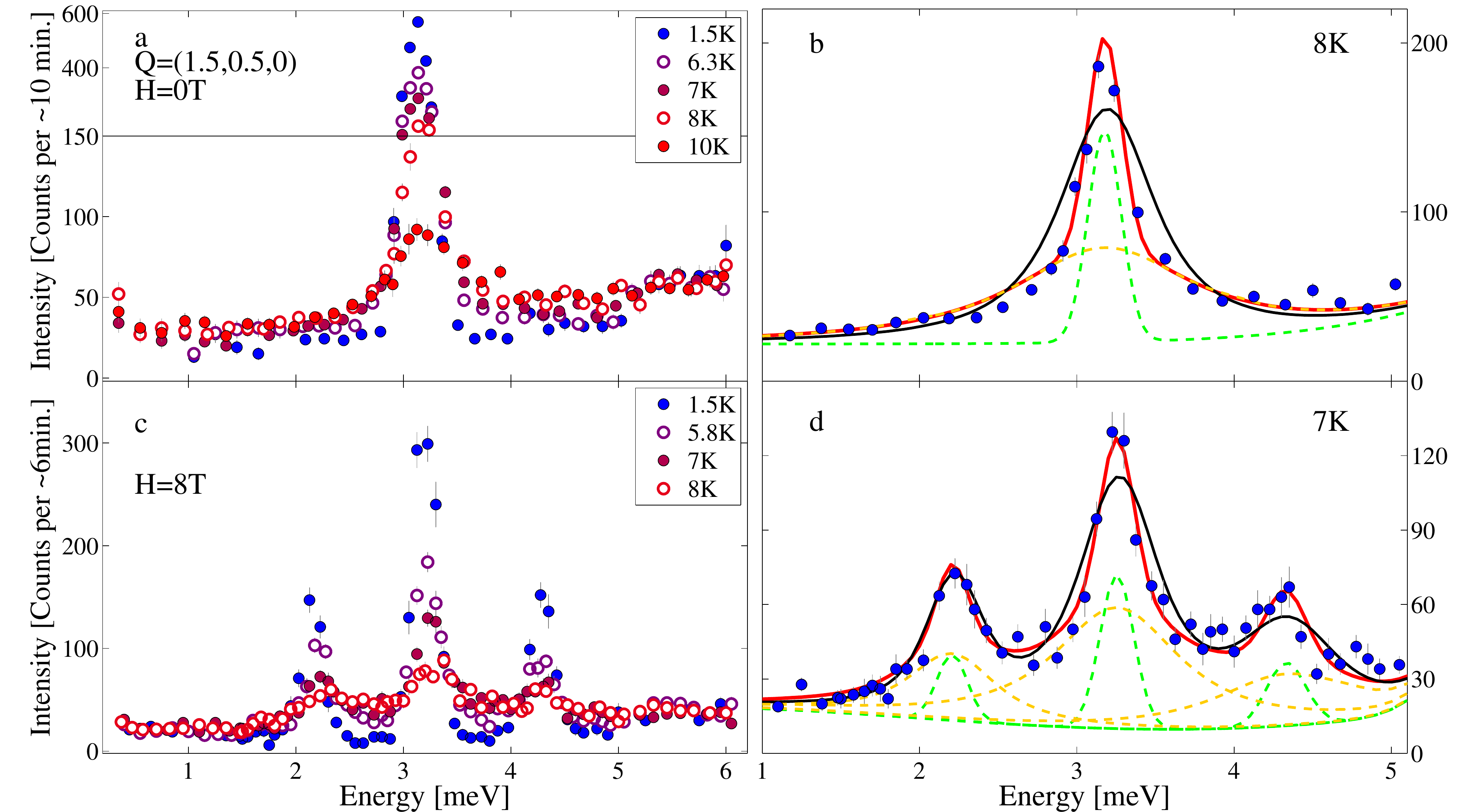}
\caption{(color online) INS spectra at Q=(1.5,0.5,0) measured on a single crystal sample.
(a) H=0~T. Measured on the TASP spectrometer at SINQ.
A two-stage intensity scale is used to visualize both the peak and the tails of the signal.
(b) Data of (a)  at $T=$8~K with fits: red line is the total fit function consisting of sharp component in green and broad component in orange. 
The black line is a fit to a single DHO which fails to match the data. 
(c) H=8~T. Measured on the PANDA spectrometer at FRM-2 with H$\parallel$c-axis.
(d) H=8~T at $T=$7~K, lines as in (b). The fitting procedure is described in the text, and the DHO is convolved with the instrument resolution.}
\end{figure}

To further investigate the damping process, measurements on single crystal were performed as a function of momentum and temperature in zero magnetic field (ZF) (Fig.~2a,b) and in applied magnetic field (Fig.~2c,d). 
We use a damped
harmonic oscillator (DHO) line-shape to fit the spectra. The neutron scattering intensity is then proportional to the dynamical structure factor given by:
%\begin{equation}
\begin{multline} 
S(Q, \omega) = I_{Q,T}  \\ \times \frac{[n(\omega,T)+1] 4 \omega
\Gamma_{Q,T}/\pi}{(\omega^2-(\omega_Q+\delta\omega_Q(T))^2-\Gamma_{Q,T}^2)^2+4\omega^2\Gamma_{Q,T}^2}
\end{multline} 
%\end{equation}
%where $\omega_{Q,0}$ is the base temperature dispersion, and  $\delta\omega_{Q,T}$ its temperature dependent shift. 
where $\omega_Q$ is the low temperature dispersion, $\delta\omega_Q(T)$ its temperature dependent shift, $\Gamma_{Q,T}$  the damping parameter (Lorentzian line-width), and $n(\omega,T)=({e^{\omega/{k_B T}}-1})^{-1}$ the detailed balance factor. The DHO integrated intensity $I_{Q,T}$ corresponding  to the spectral weight, the dispersion $\omega_Q(T)=\omega_Q+\delta\omega_Q(T)$, and $\Gamma_{Q,T}$ are the only fitted parameters.
When fitting data, this line-shape was convolved with the Gaussian energy resolution of the instrument.\\

At Q=(2,0,0) and Q=(1.5,0,0) previous high resolution INS measurements have shown that at ZF the triplet mode is slightly split into three
respectively two branches by DM interactions~\cite{Gaulin04_s,Kakurai05_s}. Data and fits for these Q values are presented in supplemental material. For Q=(1.5,0.5,0)
the mode consists essentially of a single peak, while the bound state has reduced intensity and lies at higher energy~\cite{Kakurai05_s,Aso05_s,Knetter04},
enabling a clean analysis of the single particle spectra (Figs.\ 2a,b). The application of an 8~T magnetic field splits the modes by about 1~meV and thereby allows to study of the damping for the individual $S_z =0,\pm1$ triplet modes (Figs.\ 2c,d). 

At the intermediate temperatures of $T=$5-9~K a line-shape consisting of one DHO per triplet mode fails to describe the spectra (solid black lines in Figs.\ 2b,d). For temepratures $T\ll\Delta$ simple models for Lorentzian damping have been used to fit spectra in other gapped systems \cite{Ruegg05,Zheludev08}. We thus implement a two-component damping model for each Zeeman split triplet consisting of the sum of a sharp and a broad DHO.  It turns out, over the studied temperature range, that the sharp component remains resolution limited $\Gamma^{S} = 0$. The total spectral weight (the sum of the broad and sharp components) matches within error bars the isolated singlet population, in agreement with the results from the powder sample (Fig.\ 1f). The ratios of sharp to broad components for each triplet mode are found to be similar and were therefore correlated in the fits.  The energy shifts $\delta\omega_Q$ are also similar and therefore kept correlated while the $\Gamma$ for each mode are not. After fitting the low temperature spectrum, the remaining temperature dependent parameters thus reduce to: \{$I^{S1}$, $\delta\omega^{S1}$, $\Gamma^{B1}$,
$\Gamma^{B2}$, $\Gamma^{B3}$\}, where the superscripts S and B refer to sharp and broad components and the
numbers 1, 2, 3 to the $S_z =0,\pm1$ triplet modes respectively.

At ZF, the same procedure is applied and each DM split mode (only one mode for  Q=(1.5,0.5,0)) is fitted by two DHOs, one remaining resolution sharp. The fit correlations are as before except for the $\Gamma$s which are this time also correlated as the triplets are on the same energy scale. The temperature dependent parameters are thus given by: \{$I^{S1}$, $\delta\omega^{S1}$, $\Gamma^{B1}$\} where the number 1 refers to the first DM split mode. This model produces a good agreement with the data over the entire measured temperature range (2-10~K), for all fields, and for all Q values, even though these have different intrinsic triplet structures. 

The fit results are summarized in Fig.~3. As a function of temperature, the excitations  shift to slightly higher energies as observed in the powder sample. The spectral weight of the sharp component quickly drops to zero while that of the broad one increases. With increasing temperature, there is an exchange of spectral weight between the two components with a cross-over at $T=$6.8~K in ZF ($T=$5.5~K at $H=$8~T). Around $T=$10~K in ZF ($T=$8~K at $H=$8~T), the sharp component has almost completely vanished. At 8~T, the damping widths $\Gamma$ of the upper and middle triplets are similar while the lowest is sharper indicating slightly less damping. In ZF the results obtained for 3 different Q values are in close agreement (Fig. 3a-c). 

\begin{figure}[h]
\centering
\includegraphics[width=\columnwidth]{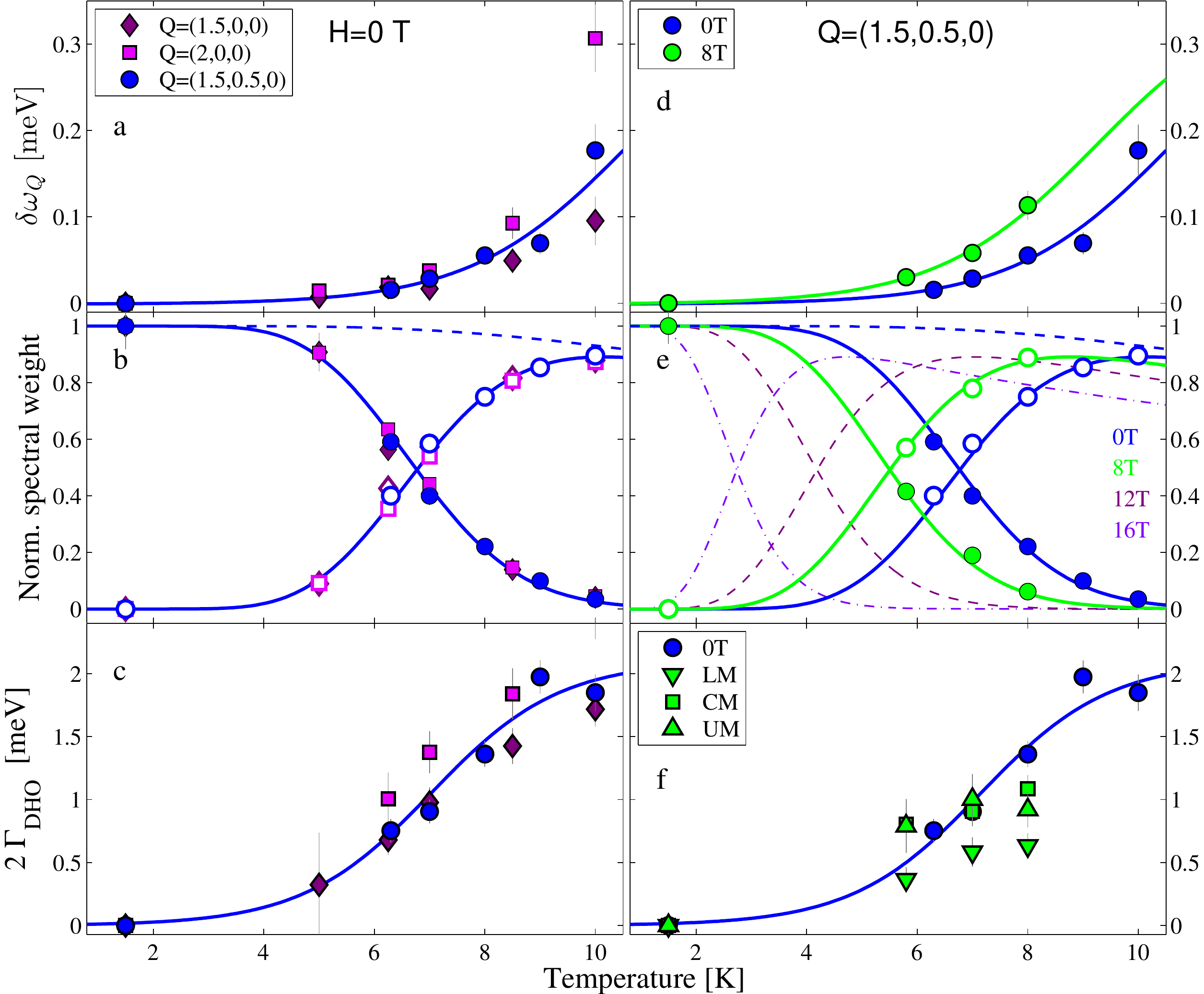}
\caption{(color online) Fit
parameters: line shift $\delta\omega_Q$ (a, d), spectral weights $I$ (b, e), and damping parameter $\Gamma$ (c, f) as a function of temperature for several momenta (left) and magnetic fields (right). In (b, e), open (closed)
symbols refer to the broad (sharp) components. The blue dashed line is the isolated singlet population at zero field. Full lines are the probabilities $P_s(T)$ and $P_b(T)$ discussed in the text, additional predictions are given for 12~T and 16~T, (g=2.22, $\Delta$=35~K). Lines in (a,c,d,f) are guides to the eye. In (f) LM, CM, UM, are the lower, middle, and upper triplet modes
respectively.}
\end{figure}

We now discuss the unconventional spectral line-shape in SrCu$_2$(BO$_3$)$_2$, which thus corresponds to a damping mechanism for the triplons with two components. The first one involves essentially undamped
triplets, that remain with a well defined (sharp) energy. This type
of triplet excitation is rapidly suppressed when increasing
temperature, and even more rapidly in the presence of a magnetic
field. The second type of triplet excitation (broad) decays quickly
with characteristic damping widths increasing with temperature. The spectral weights of these two components add up to the isolated singlet
population expected from the value of the gap and statistics. 

These results imply that the presence of a small percentage
of thermally excited triplets (2\% at 6.8~K, 9\% at 10~K) strongly modifies the thermodynamic
properties in \SCBOO. For highly frustrated systems, such behavior could reflect possible proximity of other low-lying states that are nearly degenerate with the ground state. Here we argue that due to the special localized nature of the triplets, this unusual behavior may be explained in a real space picture. Theoretical results~\cite{Shawish06,Capponi09} indicate that the introduction of a spin vacancy on one Cu$^{2+}$ site creates a sizable polarization on 6 of the neighboring dimers. In a similar manner the introduction of one triplet on the lattice polarizes 6 of the neighboring dimers, creating an extended  triplon, following a pattern calculated by perturbative continuous unitary transformations (PCUT) method~\cite{Dorier08} and illustrated in Fig.~ 4. The 6 polarized neighboring dimers correspond to an effective triplon radius of 1.27 lattice units.

Applied to our data, we propose a very simple model of correlated decay to explain the observed spectral line-shape.
Decay occurs when two triplons overlap spatially, no decay occurs when there is no spatial overlap. We consider for calculation one triplon extending isotropically over a radius $R$  on the 2D Shastry-Sutherland lattice with two dimers per unit cell and lattice constant $a$.  The temperature dependent probabilities for the sharp $P_s(T)$ and broad $P_b(T)$ components can then be computed in terms of triplon overlap: there will be no overlap if only singlets are present in the area $A=\pi (2R)^2$ extending over a radius $2R$ surrounding the triplet excited by the neutron.
\begin{equation}
P_s(T) = n_s(T)^{2\pi (2R/a)^2}.
\end{equation}
$ P_b(T)$ corresponds to having a singlet on the central dimer -- so that it can be excited by the neutron -- and at
least one triplet inside the area $A$ :
\begin{equation}
P_b(T) = n_s(T) - n_s(T)^{2\pi (2R/a)^2} =n_s(T) -P_s(T).
\end{equation}

This phenomenological model matches the experimental temperature
dependence of the broad and sharp components very well (full lines in Fig. 3 b) for a triplon radius $R=1.3a$ (11.7~\AA~for \SCBO with lattice parameter $a$=8.99~\AA). This value also remarkably matches the 8~T data (green lines in Fig. 3 e)
without any further adjustment, the shift to lower temperatures
being \emph{precisely} accounted for just by the different value $n_s(T)$ takes in the
presence of the field, as the gap reduces by $g\mu_BH$. The value we obtain for the triplon radius $R$ is in perfect agreement with the value calculated independently by the PCUT method. 
\begin{figure}[h]
\centering
\includegraphics[width=6cm]{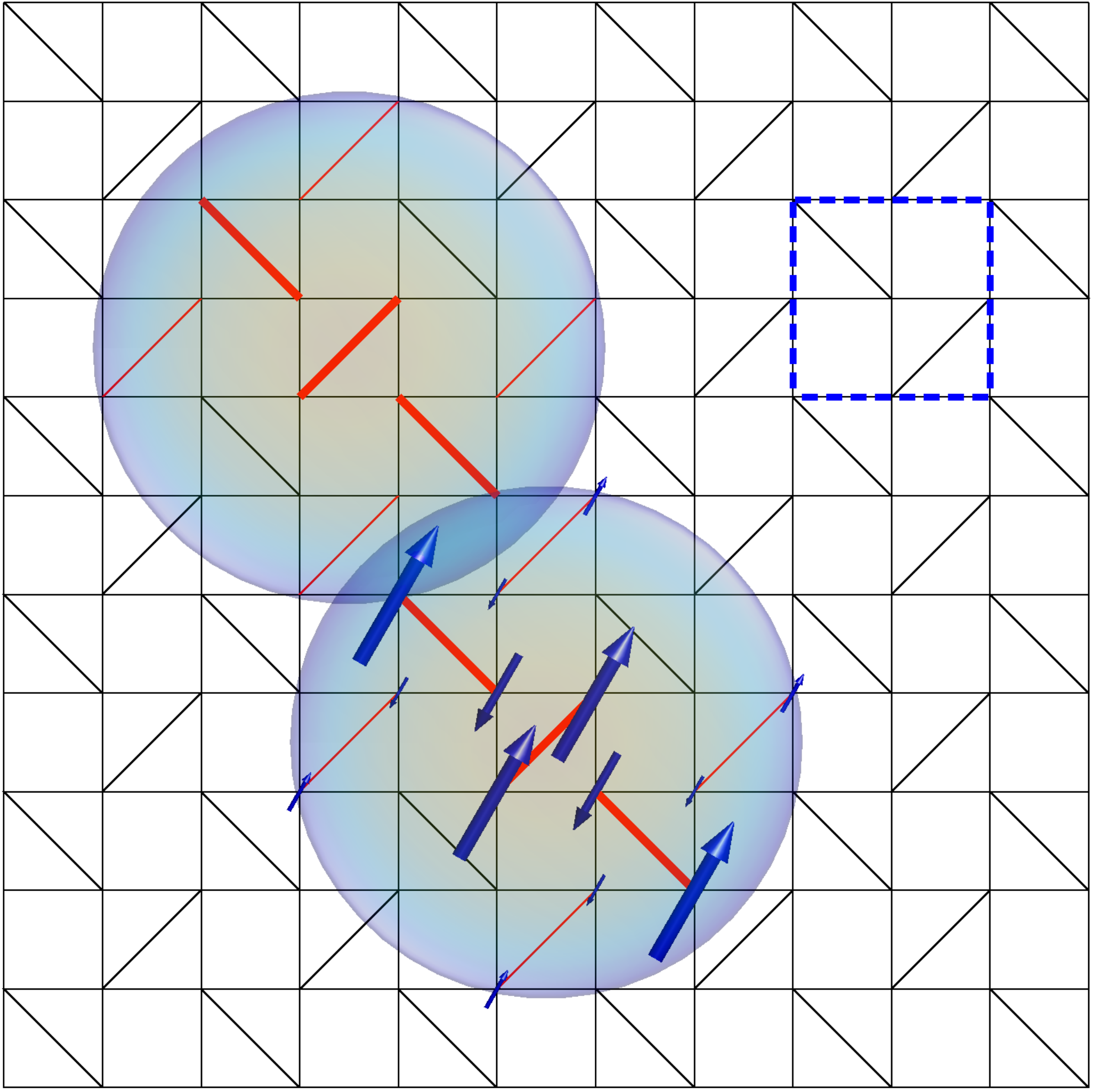}
\caption{(color online) Real space sketch of two overlapping triplons (blue disks) with radius 1.3 lattice units on the \SHS lattice. The magnetization pattern around the central triplet is shown by up down arrows with lengths proportional to the magnetization~\cite{Dorier08,Mila11_priv}. The dashed blue square indicates a unit cell.}
\end{figure}

We finish by comparing our results to observations of non-Lorentzian damping with asymmetric spectral functions that have been reported recently for some gapped low--dimensional dimer systems like the alternating chain Cu(NO$_{3}$)$_{2} \cdot$ 2.5H$_{2}$O  \cite{Tennant03,Tennant12} or the layered dimer material Sr$_3$Cr$_2$O$_8$ \cite{Castro12,Jensen14}. While these systems show finite lifetime of their quasi-particles at finite temperatures due to scattering on thermally excited triplets and hence an effect that can be understood by a density-of-states argument, the spectral line-shapes reported here for the frustrated Shastry-Sutherland compound SrCu$_2$(BO$_3$)$_2$ appear to be very different with two components and over all pagoda shape, that is very distinct from (asymmetric) Lorentzian models proposed normally for T $\ll\Delta$ \cite{Essler08}.  

In summary, we have reported detailed INS data on the temperature dependence of the magnetic excitation spectrum in the \SHS compound \SCBOO. We identified an anomalous spectral line-shape consisting of two components, one is long-lived and upon increasing temperature it transfers spectral weight to a broad, rapidly decaying component. We explain this behavior as correlated decay of triplons and establish a parameter free model of the temperature and magnetic field dependence of the cross-over between sharp and broad components.
This simple model should serve as guidance for future more accurate theoretical treatment of this fascinating damping phenomenon.

We acknowledge F. Mila, B. Normand and A. M. L\"auchli for many useful discussions. This work is based on experiments performed at the Swiss spallation neutron source SINQ, Paul Scherrer Institute, Villigen, Switzerland and at the FRM-2, Munich, Germany, and was supported by the Swiss National Science Foundation.

%\bibliographystyle{apsrev4-1}
%\bibliography{../../SCBO/Thesis_Mohamed_Zayed/ma_bib_5}
%\end{document}

%

\end{document}